\documentclass[a4paper,11pt]{article}
\pdfoutput=1 

\usepackage{jinstpub} 

\title{The environmental monitoring system at the COSINE-100 experiment}

\author[i]{H.~Kim}
\author[p]{G.~Adhikari}
\author[b]{E.~Barbosa~de~Souza}
\author[c]{N.~Carlin}
\author[d]{J.~J.~Choi}
\author[d]{S.~Choi}
\author[e]{M.~Djamal}
\author[f]{A.~C.~Ezeribe}
\author[c]{L.~E.~Fran{\c c}a}
\author[g,1]{C.~Ha,\note{Corresponding author.}}
\author[h,q,l]{I.~S.~Hahn}
\author[i]{E.~J.~Jeon}
\author[b]{J.~H.~Jo}
\author[d]{H.~W.~Joo}
\author[i]{W.~G.~Kang}
\author[j]{M.~Kauer}
\author[k]{H.~J.~Kim}
\author[i]{K.~W.~Kim}
\author[i]{S.~H.~Kim}
\author[d]{S.~K.~Kim}
\author[l,i]{W.~K.~Kim}
\author[i,a,l]{Y.~D.~Kim}
\author[i,m,l]{Y.~H.~Kim}
\author[i]{Y.~J.~Ko}
\author[i]{E.~K.~Lee}
\author[l,i]{H.~Lee}
\author[i,l]{H.~S.~Lee}
\author[i]{H.~Y.~Lee}
\author[i]{I.~S.~Lee}
\author[i]{J.~Lee}
\author[k]{J.~Y.~Lee}
\author[i,l]{M.~H.~Lee}
\author[l,i]{S.~H.~Lee}
\author[d]{S.~M.~Lee}
\author[i]{D.~S.~Leonard}
\author[c]{B.~B.~Manzato}
\author[b]{R.~H.~Maruyama}
\author[f]{R.~J.~Neal}
\author[i]{S.~L.~Olsen}
\author[l,i]{B.~J.~Park}
\author[n]{H.~K.~Park}
\author[m]{H.~S.~Park}
\author[i]{K.~S.~Park}
\author[c]{R.~L.~C.~Pitta}
\author[i]{H.~Prihtiadi}
\author[i]{S.~J.~Ra}
\author[o]{C.~Rott}
\author[i]{K.~A.~Shin}
\author[f]{A.~Scarff}
\author[f]{N.~J.~C.~Spooner}
\author[b]{W.~G.~Thompson}
\author[p]{L.~Yang}
\author[o]{G.~H.~Yu}

\affiliation[a]{Department of Physics, Sejong University, Seoul 05006, Republic of Korea}
\affiliation[b]{Department of Physics and Wright Laboratory, Yale University, New Haven, CT 06520, USA}
\affiliation[c]{Physics Institute, University of S\~{a}o Paulo, 05508-090, S\~{a}o Paulo, Brazil}
\affiliation[d]{Department of Physics and Astronomy, Seoul National University, Seoul 08826, Republic of Korea}
\affiliation[e]{Department of Physics, Bandung Institute of Technology, Bandung 40132, Indonesia}
\affiliation[f]{Department of Physics and Astronomy, University of Sheffield, Sheffield S3 7RH, United Kingdom}
\affiliation[g]{Department of Physics, Chung-Ang University, Seoul 06973, Republic of Korea}
\affiliation[h]{Department of Science Education, Ewha Womans University, Seoul 03760, Republic of Korea} 
\affiliation[i]{Center for Underground Physics, Institute for Basic Science (IBS), Daejeon 34126, Republic of Korea}
\affiliation[j]{Department of Physics and Wisconsin IceCube Particle Astrophysics Center, University of Wisconsin-Madison, Madison, WI 53706, USA}
\affiliation[k]{Department of Physics, Kyungpook National University, Daegu 41566, Republic of Korea}
\affiliation[l]{IBS School, University of Science and Technology (UST), Daejeon 34113, Republic of Korea}
\affiliation[m]{Korea Research Institute of Standards and Science, Daejeon 34113, Republic of Korea}
\affiliation[n]{Department of Accelerator Science, Korea University, Sejong 30019, Republic of Korea}
\affiliation[o]{Department of Physics, Sungkyunkwan University, Suwon 16419, Republic of Korea}
\affiliation[p]{Department of Physics, University of California San Diego, La Jolla, CA 92093, USA}
\affiliation[q]{Center for Exotic Nuclear Studies, Institute for Basic Science (IBS), Daejeon 34126, Republic of Korea}
\collaboration{COSINE-100 Collaboration}

\emailAdd{chha@cau.ac.kr}

\abstract{
  The COSINE-100 experiment is designed to test the DAMA experiment
  which claimed an observation of a dark matter signal from an annual modulation in their residual event rate.
  To measure the 1~\%-level signal amplitude, it is crucial to control and monitor nearly all environmental quantities
  that might systematically mimic the signal.
  The environmental monitoring also helps ensure a stable operation of the experiment.
  Here, we describe the design and performance of the centralized environmental monitoring system
  for the COSINE-100 experiment.
}


\arxivnumber{2107.07655}

\begin{document}
\maketitle
\flushbottom

\section{Introduction}

There is indirect evidence for the existence of dark matter in the universe~\cite{Clowe:2006eq,Huterer2010,Ade:2015xua}.
However, it has not been detected in a laboratory-based experiment, with the notable exception of
the long-debated claim by the DAMA/LIBRA collaboration~\cite{Bernabei:2013xsa,Bernabei:2018yyw}.
Since 2003, DAMA has been reporting an annual modulation signal in their residual event rate~\cite{Bernabei:2003za}
that is about 1\,\% in amplitude over background with a one year period as might be expected from Weakly Interacting Massive Particle (WIMP)
dark matter interactions in their NaI(Tl) target crystals~\cite{Freese:2012xd}.
In order to explain the signal, many new ideas and models have been proposed and the result remains a great mystery
~\cite{Kopp:2009et,Davis:2014cja,Nygren:2011,Ralston:2010bd,Blum:2011jf,PhysRevD.93.115037,Savage:2008er}.
The suggested causes for the modulation are temperature variations, radon-level variations, and cosmic-ray muons and their induced neutrons.

The COSINE-100 experiment, an underground dark matter detector at the Yangyang underground laboratory(Y2L) in Korea,
is designed to test this claim using the same target medium~\cite{Adhikari:2017esn}.
To achieve the 1\,\% level sensitivity, ideally, all the environmental variables should be controlled better than this level;
if not they should be carefully monitored so that correlation studies of environmental data with dark matter search data
can be performed.

The Y2L facility is situated next to the Yangyang pumped-storage hydroelectric power plant
that is located at a depth of 700~m.
The experimental tunnel area is composed of gneiss rocks that provide an overburden of 1800 meter-water-equivalent (m.w.e) depth.
A 2~km-long drive--way allows access to the laboratories as well
as air ventilation to the facility.
In the location for the Korea Invisible Mass Search (KIMS) experiment, we measured 44.4$\pm$18.1 Bq/m$^3$ for radon,
23.0$\pm$1.0$^{\circ}$C in temperature, and 23$\pm$1\,\% in relative humidity~\cite{Lee:2011jkps}.
Stable power is provided to the experimental areas by staged uninterruptible power supplies (UPS).

For stable data-taking and systematic analyses of the seasonal variations, it is essential to monitor
environmental parameters such as detector temperatures, room humidities, high voltages and currents to light sensors, cosmic-ray induced effects,
and DAQ stability. Therefore, we employ various monitoring devices controlled
by a common database server and an integrated visualization program.
In addition to checking the experimental environment,
the monitoring also provides safety measures in case of an emergency in the tunnel.
As the number of monitoring devices grows, it is important to have a robust design
that integrates various devices readily.
In this article, we describe the slow monitoring system for the COSINE-100 experiment.

\section{The COSINE-100 experiment}
The COSINE-100 experiment is installed at a newly developed tunnel at the Y2L facility.
The COSINE-100 detector room at the tunnel has a 44~m$^2$ floor area and is managed as a clean air environment.
The detector room is equipped with several environmental monitoring systems
to control humidities, radon level, and temperatures.
Automatically regulated electrical power through uninterruptible power supplies is provided,
and the stability of the power is continuously monitored, although blackouts rarely happen.

\begin{figure*}[!htb]
  \begin{center}
    \begin{tabular}{cc}
      \includegraphics[width=0.66\textwidth]{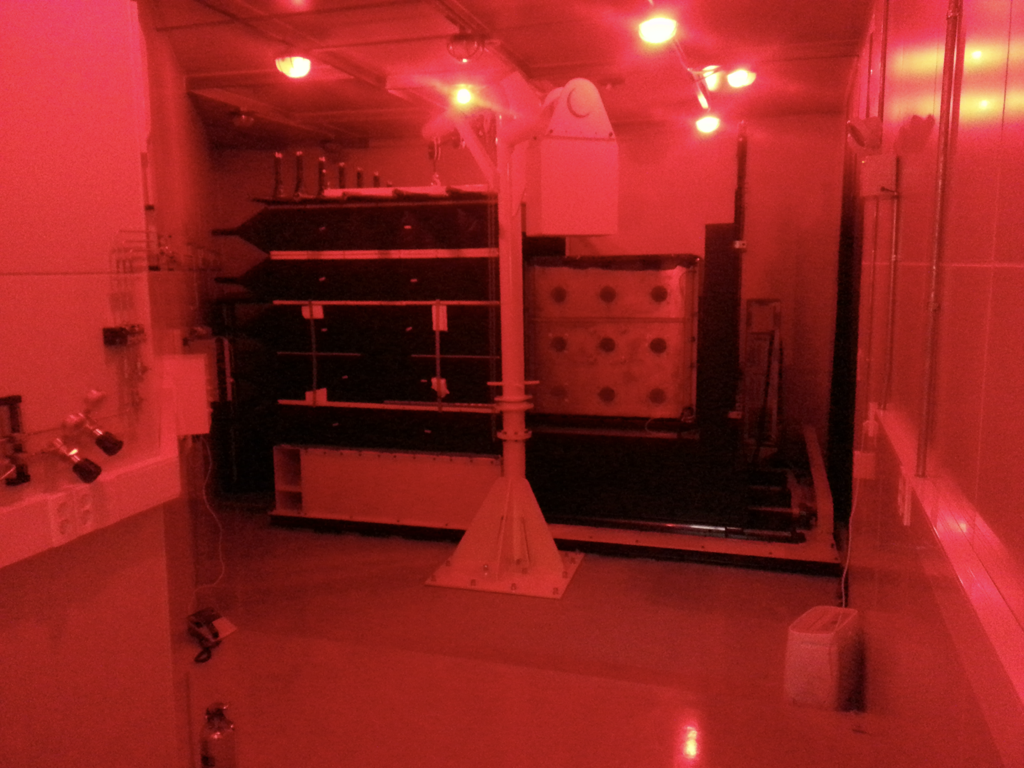} &
      \includegraphics[width=0.28\textwidth, angle=0]{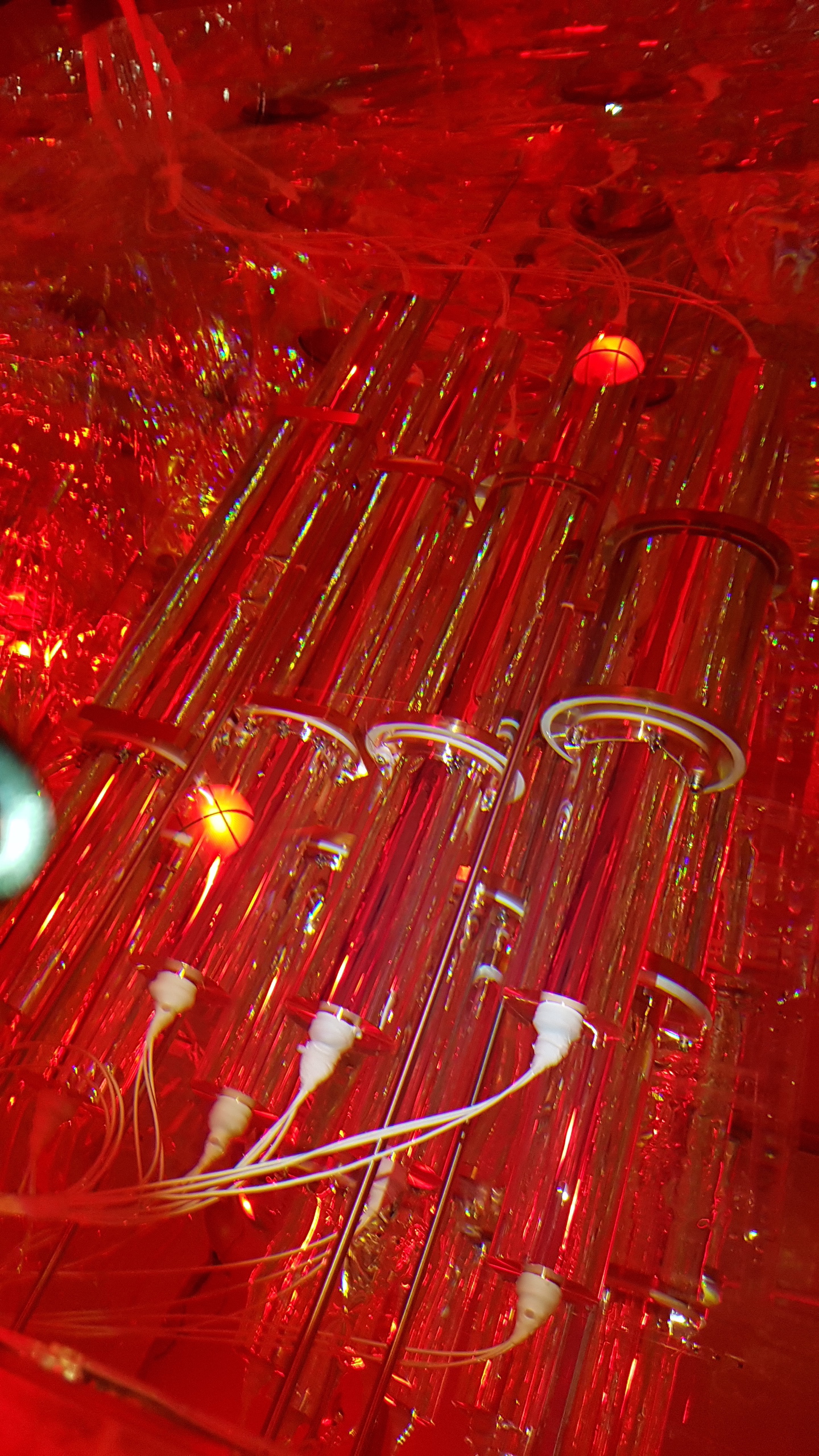} 
    \end{tabular}
  \end{center}
  \caption{Pictures of the COSINE-100 detector room (left) and the crystal array in the liquid scintillator (right). }
  \label{ref:crystals}
\end{figure*}

Eight NaI(Tl) crystal detector~\cite{bkgunderstanding1,bkgunderstanding2} assemblies are arranged in a 4$\times$2 array and placed
on an acrylic table in the center of the main experimental structure.
The total crystal mass is 106~kg. These crystal detectors are light-coupled to 3-inch Hamamatsu R12669SEL PMTs~(selected for high quantum efficiency).
To attenuate or tag externally generated radiation from cosmogenic radioisotopes, and intrinsic radioactivity
in the environment and various external detector components,
the detector is surrounded by layers of passive and active shielding materials.
From the center outwards, the four shielding layers include 2200-liter scintillating liquid~\cite{ls}, copper box, lead-brick walls, and plastic scintillator panels~\cite{mu1,mu2}.
Since the crystal array is submerged in a large volume of scintillating liquid
that has a relatively high heat capacity,
the expected temperature variations near the crystals are less than $\pm~0.1^{\circ}$C.
The COSINE-100 experiment started the data-taking in September of 2016
and runs continuously with more than 90\,\% livetime so far, with which
several interesting results have been produced~\cite{result1,result2,result3}.
Pictures of the COSINE-100 detector room and crystals in the liquid scintillator
are shown in Fig.~\ref{ref:crystals},
and further details of the experimental setup can be found elsewhere~\cite{Adhikari:2017esn}.

\section{Design of the environmental monitoring}

\begin{figure}[htbp]
\begin{center}
\includegraphics[width=0.9\textwidth]{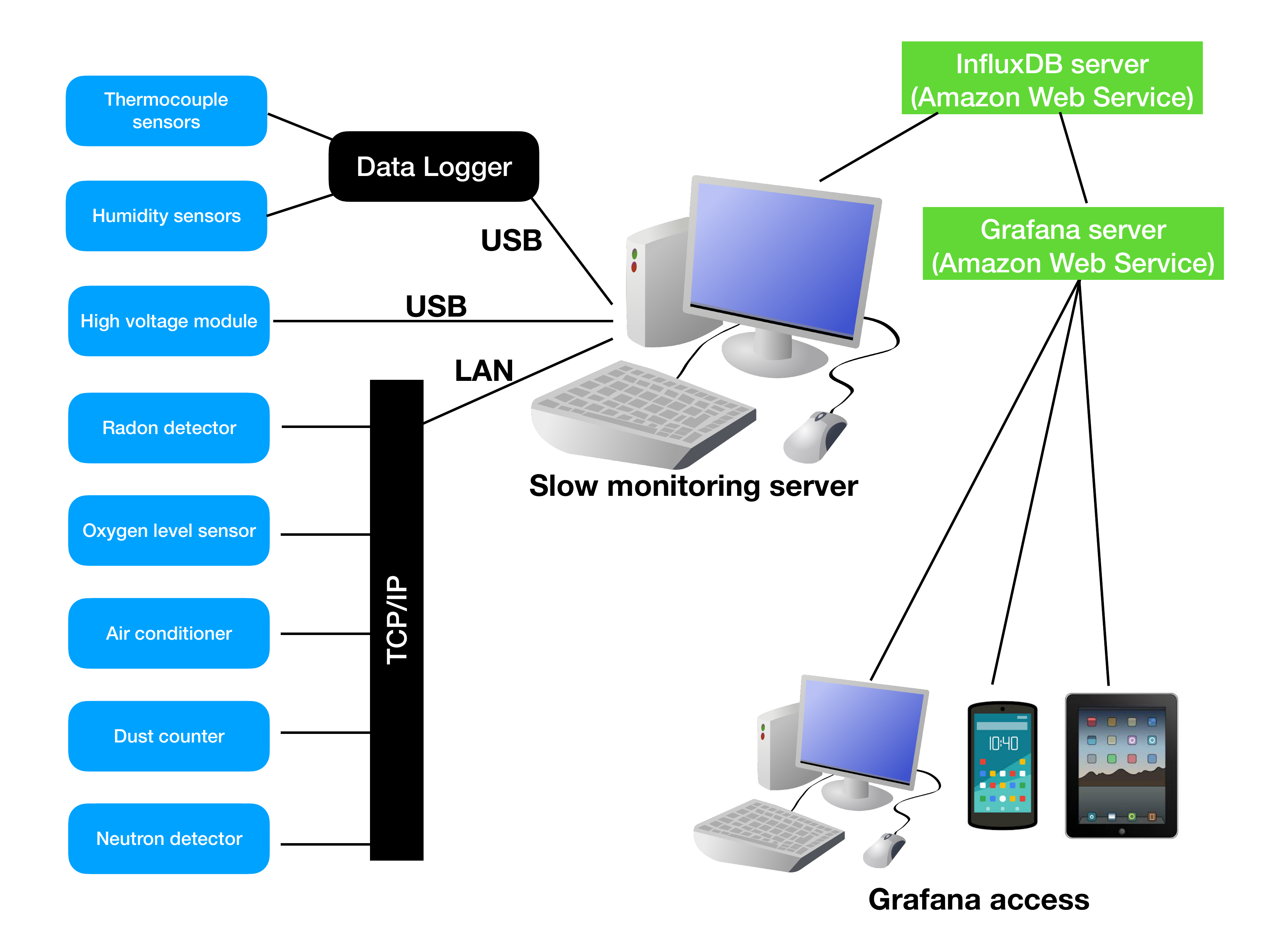}
\caption{A schematic of the environmental monitoring data flow. Monitoring parameter data are transferred via hypertext transfer protocol and universal serial bus (USB) to
  a slow monitoring computer at the COSINE-100 experimental room where they are reformatted into a common structure and sent to the InfluxDB. Any display device can access the time-series data from the InfluxDB.}
\label{flow}
\end{center}
\end{figure}

The COSINE-100 environmental data are used to monitor and control the hardware.
The monitoring process involves the collection, visualization and analysis of data via an Internet connection.
The data format consists of timestamps at device-specific time intervals.
We use the InfluxDB~\footnote{www.influxdata.com. InfluxDB is a time-series database.} database for the environmental monitoring.
A schematic view of the environmental monitoring data flow is shown in Fig.~\ref{flow} and
hardware details are listed in Table~\ref{hardware}. 

The collected data are stored in two InfluxDB servers.
One is a local database server and the other is located at the Amazon web server (AWS) in Seoul.
Since the duplicate AWS database is located outside the lab,
a researcher can monitor the laboratory environment with the synchronized logs from AWS
when the connection to the laboratory server is lost in various situations.
The AWS database provides a visualization service using the Grafana application~\footnote{https://grafana.com}.
The Grafana visualization program displays sensor data plots in dynamic time
scales from five minutes to several years with parameter alert capability. 

The monitoring system consists of several independent programs that run in parallel.
These monitoring programs communicate with each other using standardized data formats,
not depending on the particular platform.
The monitoring message data are on JSON format~\footnote{https://www.json.org}
and follow the InfluxDB data structure.  

Since the underground laboratory is not easy to access overnights or on weekends,
an alert system is necessary to cope with unexpected events.
On the Grafana page, we can set alert thresholds for all the parameters
and triggered alerts are relayed to the Slack application~\footnote{www.slack.com} or an email. 

Subscribers can get an alert via a smartphone or a PC
when a parameter value is out of range relative to a set value.
In case of an unexpected event, experimental systems can be controlled via remote Internet access.
In the case of an event that occurs when the system is inaccessible, core systems, including high voltage power supplies to PMTs
and DAQ components, are programmed to be automatically shut down
after waiting for 10 minutes using a secondary UPS that can sustain operations for 20 minutes. 

\begin{table*}[!htb]
  \begin{center}
    \caption{List of hardware used in the COSINE-100 monitoring data collection.
      Specifications for the facility-provided infrastructure
      including electricity and ventilation are not listed here
      but those data are duplicated to the same database
      and monitored with the same visual program.}
    \label{hardware}
    \begin{tabular}{lcccc}
      \hline
      \hline
      Variable & hardware &  brand  & connection &  \\
      \hline
      Temperature &  K-type thermocouple  & OMEGA &  \\
                  &  TC-08 DAQ  & OMEGA &  USB\\
      \hline
      Humidity    &  MM2001  & Maxdetect &  \\
                  & U3 DAQ  & LabJack &  USB \\
      \hline
      Radon       &  RAD7 & Durridge & RS-232 Ethernet \\
      \hline
      Oxygen      &  O2H-9903SD & Lutron & RS-232 USB \\
      \hline
      Air conditioner &  A/C &Korea Air Condition Tech.& RS-485 USB \\
      \hline
      Dust level &  ApexP3 & Lighthouse & Ethernet \\
      \hline
      Local backup power & UPS  & APC & Ethernet \\
      \hline
      High voltage & SY4527 & CAEN & Ethernet \\
      \hline
      \hline
    \end{tabular}
  \end{center}
\end{table*}

\section{Monitoring parameters}

\begin{figure}[htbp]
  \begin{center}
    \includegraphics[width=0.9\textwidth]{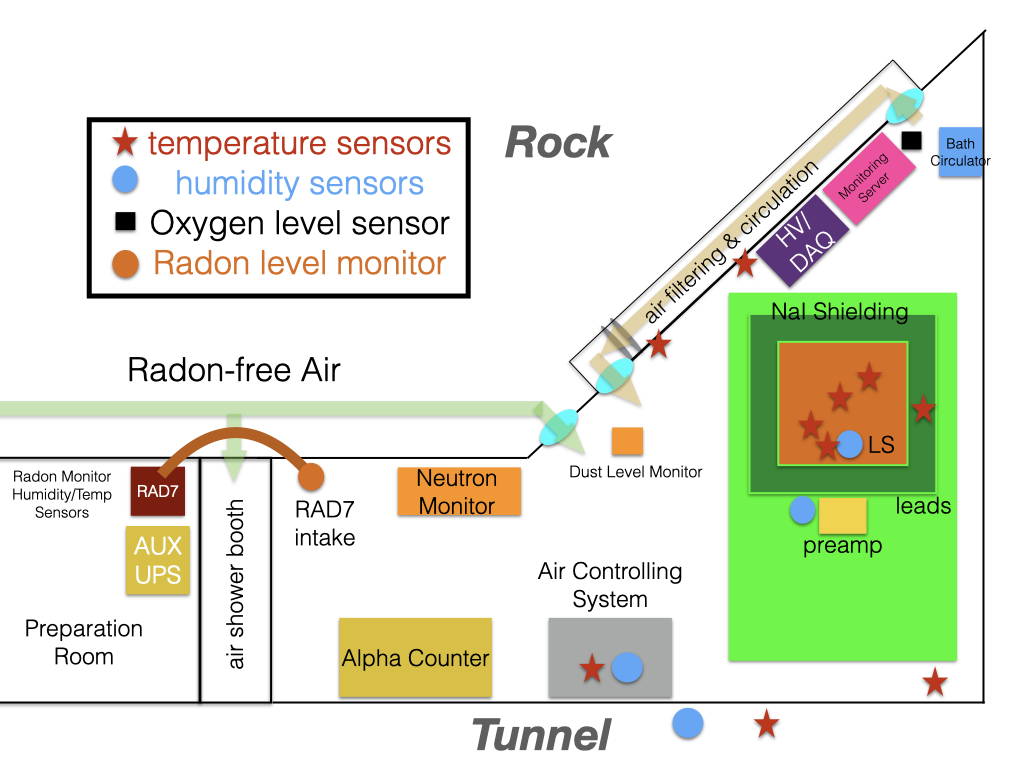}
    \caption{A floor diagram of the COSINE-100 experimental room.
      Sensor locations are marked with various shapes and colors.
    }
    \label{sensors}
  \end{center}
\end{figure}

Monitoring data from each device are collected in the monitoring server
and the time-stamped information is transferred to the InfluxDB.
The monitoring location of each device is depicted in a floor diagram as shown in Fig.~\ref{sensors}.
On-site crews maintain the devices in a regular manner
while online shift takers are responsible for the monitoring through the Grafana outputs.

\subsection{Temperature}
Characteristics of NaI(Tl) crystals are known to be temperature dependent~\cite{IANAKIEV2009432,Sailer}. 
At room temperature, the known variation of the light output is $-0.3\%/^{\circ}$C~\cite{SaintGobain}.
Therefore, to measure the 1~\%-level annual modulation effect, the temperature variation should be controlled better than $\pm$3.3$^{\circ}$C. We maintain the temperature variations near the crystals better that $\pm$0.1$^{\circ}$C. As the light output decreases with the temperature increase,
the correlation studies are important because the modulation effect gives positive amplitudes at summer seasons.

\begin{figure}[htbp]
  \begin{center}
    \includegraphics[width=0.9\textwidth]{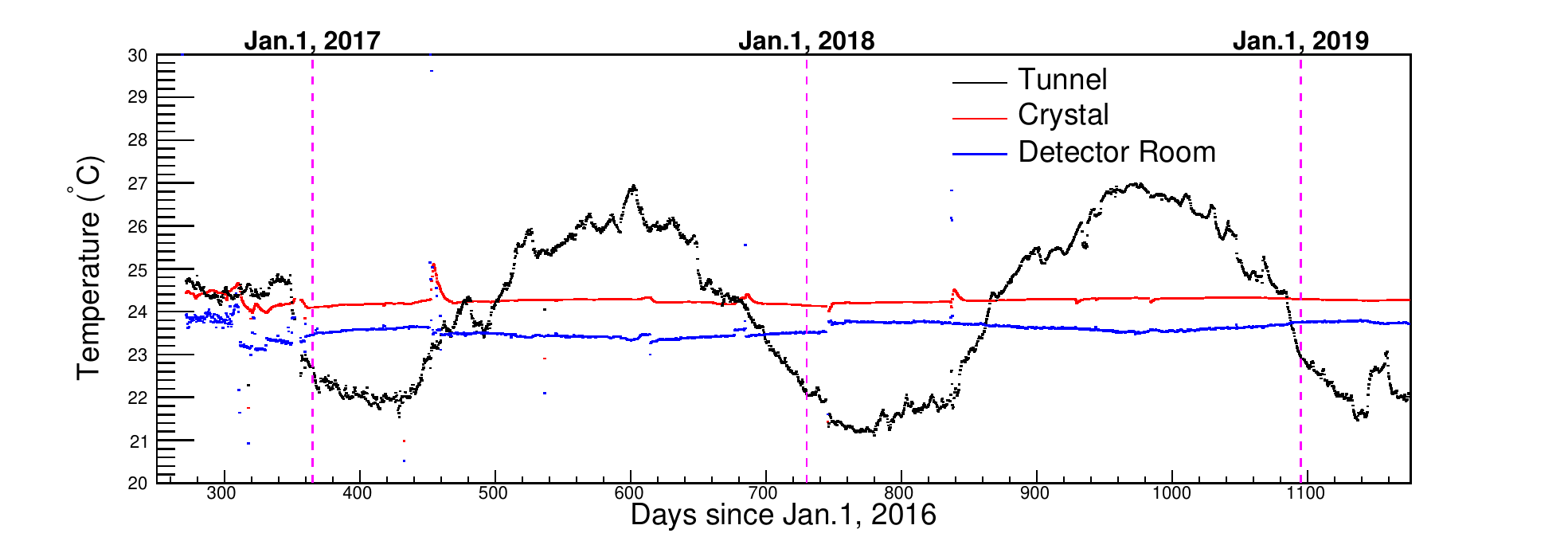}
    \caption{Long-term monitoring of temperatures in the COSINE-100 experimental area.
      Three temperature readings for a 2.5-year operation period are indicated with
      different colors. The unregulated tunnel temperature is drawn with black points,
      the temperature of the air conditioned detector room is displayed with blue points,
      and the temperature of the crystal array is indicated with red points.
      Note that occasional spikes in the crystal temperature are mainly due to power failures;
      those data are excluded in the analysis.
    }
    \label{temp}
  \end{center}
\end{figure}

To monitor temperatures at various positions in the COSINE-100 experimental area,
we use an 8-channel data logger, TC-08, from Pico Technology.
Eight K-type thermocouples~\footnote{The standard limit of errors provided by the producer is $2.2^{\circ}$C} are connected to the data logger, three of which 
are installed near the crystal detectors in contact with the liquid scintillator
and the rest are placed in the room and the tunnel.
Two separate temperature sensors in the air conditioner and the DAQ logger
cross-check the K-type thermocouple readings showing a similar or better level of
variations for month-scale measurements.
Three K-type thermocouple measurements inside the liquid scintillator at different locations
showed the same level of variations, smaller than $\pm$0.1~$^\circ$C.
The temperature data with timestamps are sent to the monitoring computer via USB.
The three temperature measurements as a function of time for about 2.5 years are shown in Fig.~\ref{temp}.

\subsection{Relative humidity}
\begin{figure}[htbp]
  \begin{center}
    \includegraphics[width=0.9\textwidth]{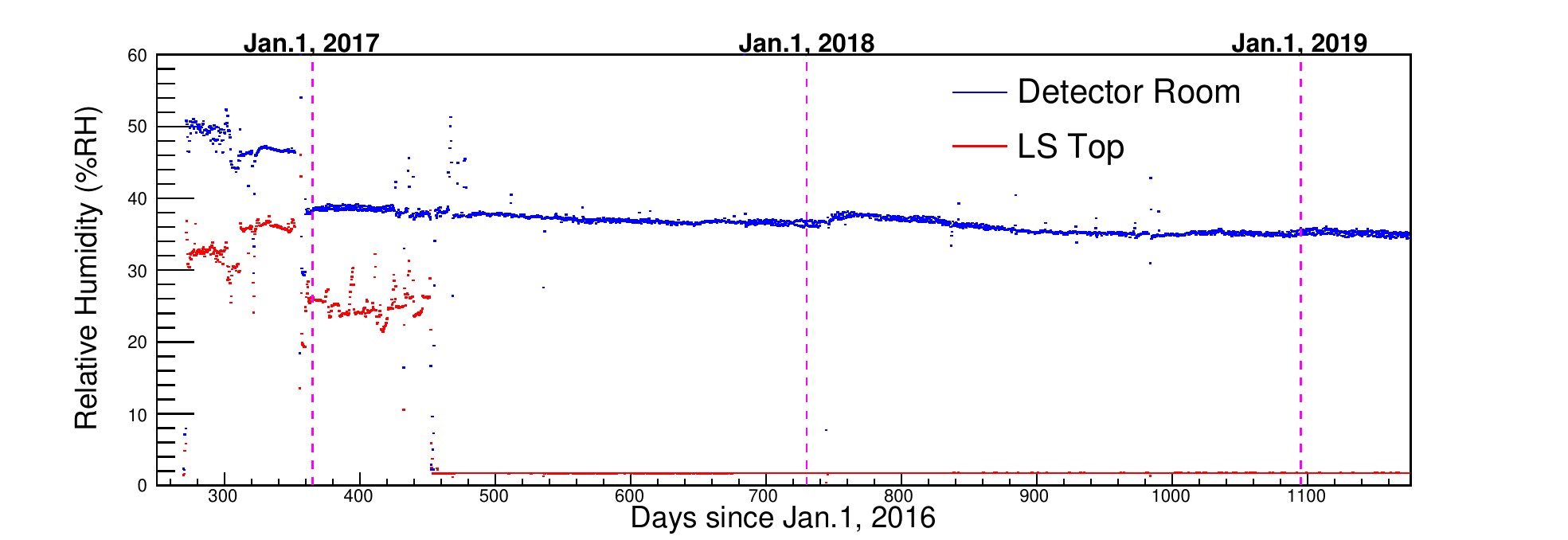}
    \caption{Long-term monitoring of humidity in the COSINE-100 experimental area.
      The panel shows two humidity readings for 2.5 years of operation.
      Blue is the detector room humidity while red indicates the humidity of the top of the liquid
      scintillator where continuous flow of N$_2$ gas is supplied. Value changes around 450 days
      are due to sensor relocation and reconfiguration. Note that the ``LS Top'' humidity measurement shows
      1.7\,\% on average which is below the specified limit of the sensor.
    }
    \label{humi}
  \end{center}
\end{figure}

Control of relative humidity is important because it can affect the 
scintillation efficiency of the liquid scintillator~\cite{LSquenching} and the stability of the DAQ system and related electronics.
When oxygen is fully removed from the liquid scintillator, the light yield is known to increase as much as 11\%.
Therefore, keeping the liquid free from oxygen would make the light yield stable. 
In the detector room, we continuously run two dehumidifiers that are set to 40\,\% relative humidity.
To reduce air contact with the liquid scintillator, we flow N$_2$ gas at about 4.0 liters per minute
over the surface of the liquid. The humidities in those areas are continuously measured.
We use three MM2001 analog sensors, manufactured by Maxdetect, and one reading from the air conditioner
to measure the relative humidity.
The humidity sensors are connected to the slow monitoring server via a Labjack U3 DAQ module.
The sensors specify a measurement accuracy of $\pm$3\,\% with the limit at 5\%~RH.
The three measurements of the humidity data as a function of time for a 2.5-year period are shown in Fig.~\ref{humi}.
We found no noticeable liquid scintillator quality change
due to temporal increase of humidity or sudden stoppage of Nitrogen gas flow throughout the measurements. 

\subsection{Radon in air}
\begin{figure}[htbp]
  \begin{center}
    \includegraphics[width=0.9\textwidth]{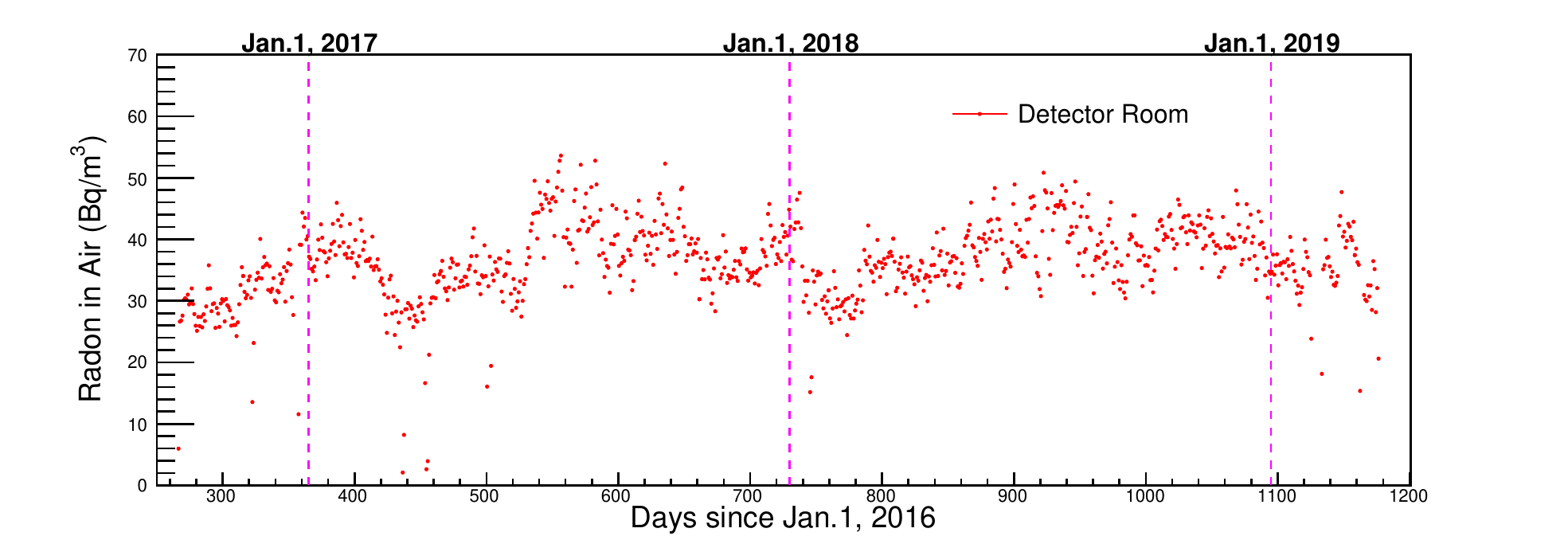}
    \caption{Long-term monitoring of radon concentration in the COSINE-100 experimental area.
      The RAD7 detector reports the radon level every 30 minutes and displays it with
      a Grafana visualization format.
    }
    \label{rad7}
  \end{center}
\end{figure}

\begin{figure}[htbp]
  \begin{center}
    \includegraphics[width=0.7\textwidth]{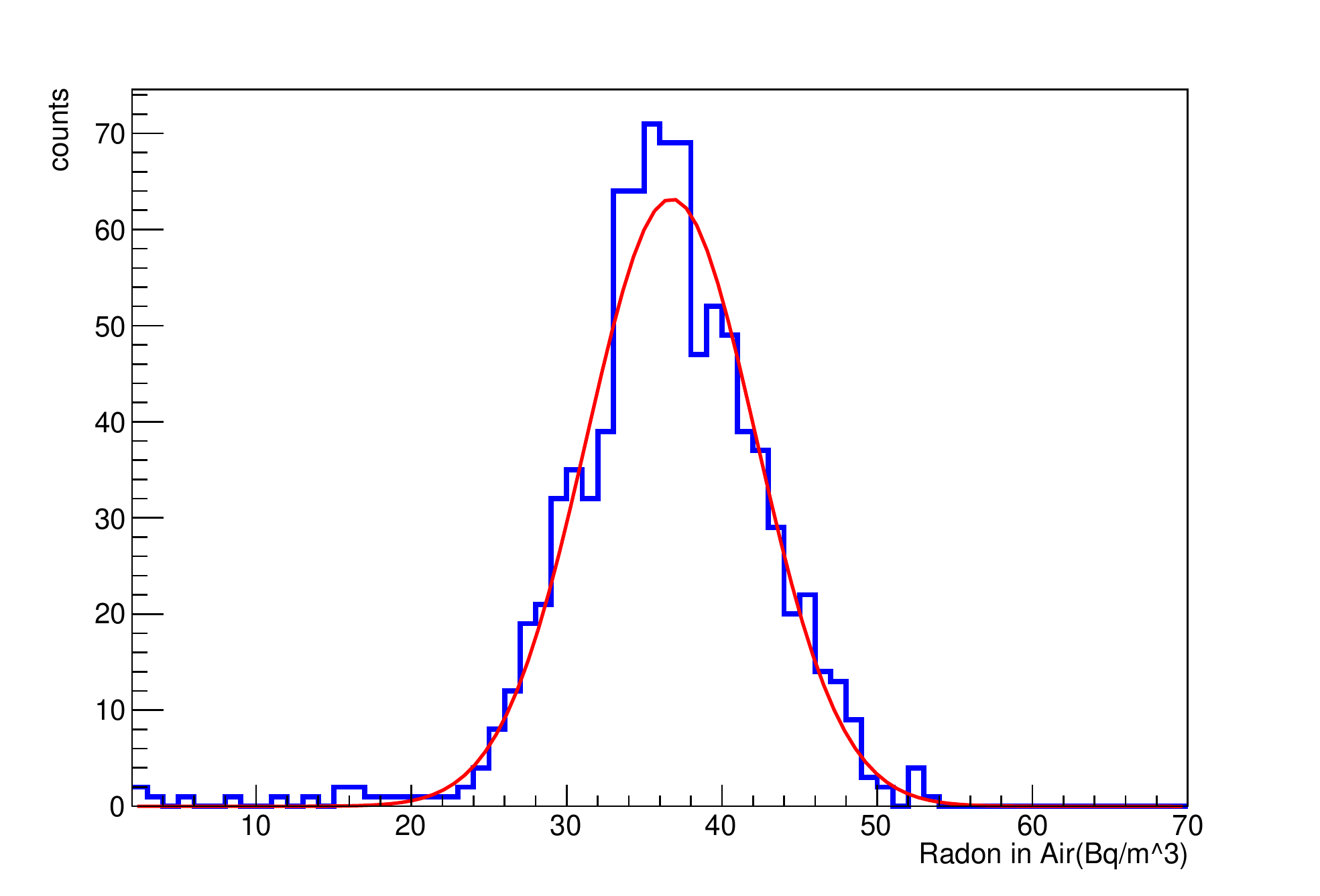}
    \caption{The variation of the radon monitoring over two and half years of data collection.
      A Gaussian fit is overlaid on the measurement.
      The fit mean and standard deviation are 36.7 and 5.5 Bq/m$^3$, respectively.
    }
    \label{radvar}
  \end{center}
\end{figure}

Radon is produced as a daughter decay product from surrounding radioactive materials in the tunnel~\cite{Lee:2011jkps}.
The radioactivity of the Y2L rock has been measured with inductively coupled plasma-mass spectrometry.
The rock samples, on average, contain 2.12~ppm of uranium and 13.02~ppm of thorium that are the primary sources of radon.
When radon decays to its daughter isotopes, several gamma radiations are produced and they can 
contribute to the background spectrum in the COSINE-100 data.
Therefore, monitoring radon in the air is important for correlation studies with the annual modulation analysis data.
We installed one RAD7 from Durridge Company for regular monitoring of the radon level in the detector
room. In RAD7, a solid-state detector is located at the center of a
drift chamber with an electric field applied.
When a $^{222}$Rn nucleus decays near the middle of the chamber, it becomes a positively charged $^{218}$Po ion
that drifts to and sticks on the detector surface.
After about 3 minutes, this $^{218}$Po subsequently decays to a $^{216}$Pb nucleus plus an alpha particle whose
alpha energy is deposited in the solid-state detector with a rate that reflects the $^{222}$Rn activity.
The total radon level is measured every 30~minutes for the COSINE-100 detector room air,
and the data are sent to the slow monitoring server that displays
the level as shown in Fig.~\ref{rad7}.
The detector specifies 10\,\% uncertainty at 150 Bq/m$^3$ level in two hours.
In the detector room, we record the radon level continuously with an average value of 36.7$\pm$0.2 Bq/m$^3$ over two and a half years
with its standard deviation of $\pm$5.5 Bq/m$^3$ as shown in Fig.~\ref{radvar}.
When the shielding has to be opened, we supply radon-free air into the detector room
that reduces the radon level down to a few Bq/m$^3$. 

\subsection{High voltage equipments}
\begin{figure}[htbp]
  \begin{center}
    \includegraphics[width=0.9\textwidth]{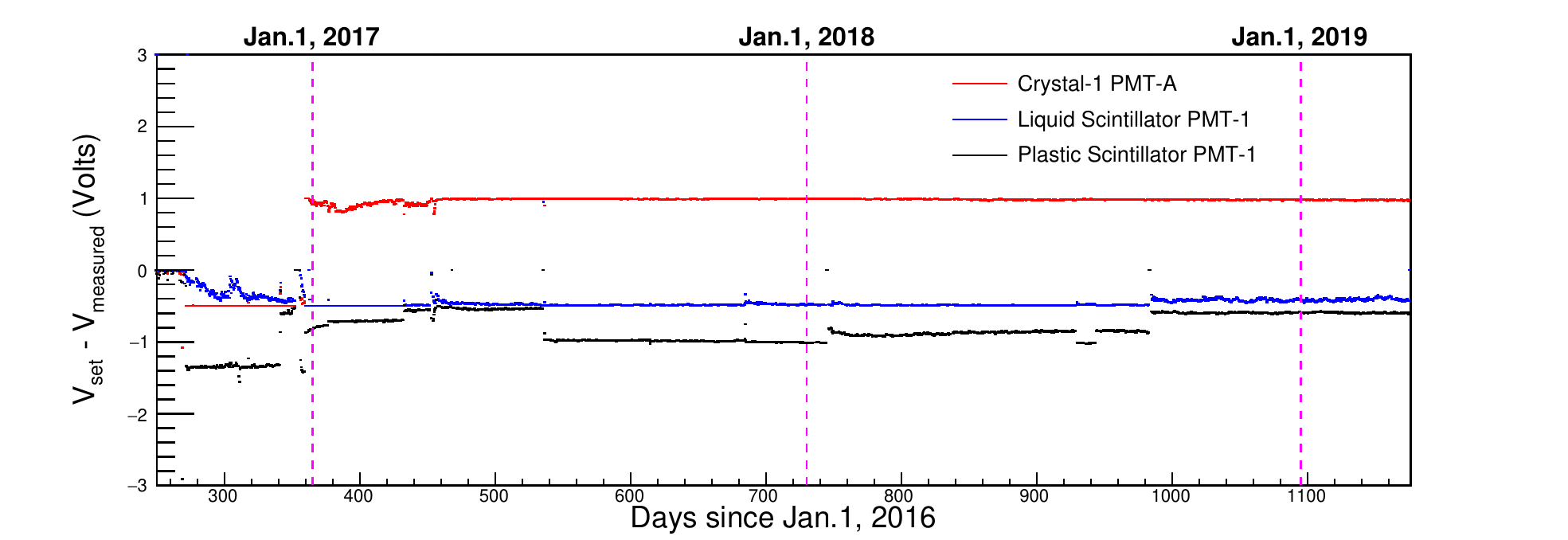}
    \caption{Long-term monitoring of high voltages of three PMTs installed in the COSINE-100 detector.
      The first PMT in each sub-detector is displayed here. The red is for one of the crystal PMTs, the blue
      is for a liquid scintillator PMT, and the black is for a plastic scintillator PMT, to which
      the supplied voltages are 1241, 1242, and 1813 volts, respectively.
    }
    \label{hv}
  \end{center}
\end{figure}

High voltages are supplied to 77 PMTs in the COSINE-100 detector system.
Since the voltage is sensitively related to the gain of a PMT,
we monitor the difference between the supplied and measured voltages, and the measured current offset.
We maintained both the voltage differences and current deviations to within 0.1\,\% for a period of more than three years.
The CAEN high voltage supply modules are connected via internal network
and their monitoring is done by a customized wrapper program.
All supplied voltages, currents, and status are recorded every minute
and sent to the monitoring server.
The long-term monitoring of the high voltage variation is shown in Fig.~\ref{hv}.

\subsection{Dust Level}

\begin{figure}[htbp]
  \begin{center}
    \includegraphics[width=0.9\textwidth]{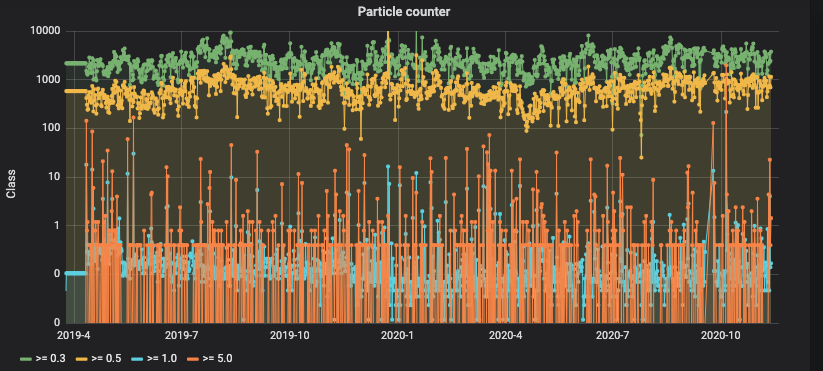}
    \caption{Dust level measurements over time displayed
      in the Grafana visualization format.
      The y-axis corresponds to the measurements of dust counts per cubic foot, also known as class.
      When measured with particulates larger than 0.5 micrometers (yellow points with lines), if the counts
      are less than 1000 per cubic foot, the experimental area
      can be classified as class 1000 cleanroom. The x-axis indicates the year-month of the measurement and
      it spans from April 2019 to December 2020. Other lines indicate different-sized dust particulates.
    }
    \label{apex}
  \end{center}
\end{figure}

A blow-filter unit that circulates and filters the detector room air runs constantly.
To monitor the dust level in the room, we installed an APEX airborne particle counter\footnote{https://www.golighthouse.com/en/airborne-particle-counters} that counts dust particulates of different sizes.
We recorded the dust level throughout the year
and found that the average dust counts over 0.5--micrometer diameter are less than 1000 counts/ft$^3$,
which meets the requirement of a class 1000 cleanroom.
The variation of the dust level does not show any time-dependent behavior (see Fig.~\ref{apex}).

\subsection{Neutron Level}
Neutrons are generated in the ($\alpha$, n) reactions by the alpha particles from the decay of $^{238}$U and $^{232}$Th, and in the spallation process by the incoming cosmic-ray muons.
When high-energy muons interact in the detector, the secondary neutrons can produce signals that mimic WIMP-nucleon interactions.
Since the high-energy cosmic-ray muon flux has a well established annual modulation,
monitoring the neutron flux is of critical importance. 

\begin{figure}[htbp]
  \begin{center}
    \includegraphics[width=0.9\textwidth]{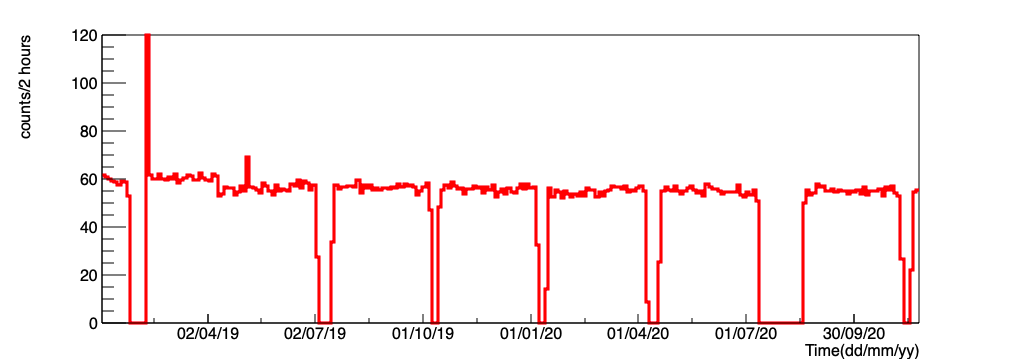}
    \caption{Thermal neutron measurements in the COSINE-100 detector room.
      The measurement was performed between January 2019 and November 2020.
      We see that the thermal neutron level is stable over the two years.
      Note that there were occasional dropouts
      due to power failures, computer reboots, and software updates. 
    }
    \label{thn}
  \end{center}
\end{figure}

We have two neutron detectors in the detector room.
One is made of DIN-based liquid scintillator for measuring fast neutrons~\cite{Adhikari:2018fmv}
and the other is a $^3$He gas detector for thermal neutrons.
The collected thermal neutron count rate is shown in Fig.~\ref{thn}.

\subsection{Other monitoring components}
Several other components that help diagnose the detector status and the room conditions
are also logged in the database.
These include DAQ computer status, pre-amplifier output voltage, oxygen monitoring,
N$_2$ flow level, air conditioner status, and backup power status.

We use a U3 DAQ module to monitor the voltage variation of the pre-amplifiers.
The PMT output signals are amplified and passed to DAQ via the custom board
which functions properly with a bias level of 6 volts. The detector room oxygen level is monitored with a Lutron O2H-9903SD device that contains an RS-232 port for serial communication. For safety purposes, we display the oxygen level data at the front of the detector room.
Korea Air Conditioning Technology provides a communication protocol based on RS-485 and Modbus to monitor the status of the air controlling system installed in the detector room.
We convert the RS-485 signal to USB for continuous monitoring.
The air conditioner data include the status of the equipment,
the temperature and humidity, and various alarms.

The APC backup power provides various parameters that can be used to check their status.
The core of the detector systems including DAQ and high voltages
is connected to this backup battery in case of both blackout and main UPS run-out.
This backup power can supply 2 Amperes for 20 minutes.
Once the backup power is triggered, an expert shift taker can log in to the DAQ system
and safely shut down the detector.
We also installed a program that automatically shuts down the DAQ system
in case of continuous battery draw for more than 10 minutes.
It is helpful when internet is also disconnected due to the power outage.

An 80~KVA online UPS protects many pieces of equipment.
The main UPS manufactured by Ewha Electronics has a network-based monitoring module that provides various protocols. Our slow monitoring system checks its status every 5 seconds via a Simple Network Management Protocol that monitors input and output voltages, as well as various event logs.

\section{Results}
\begin{figure}[htbp]
  \begin{center}
    \includegraphics[width=0.9\textwidth]{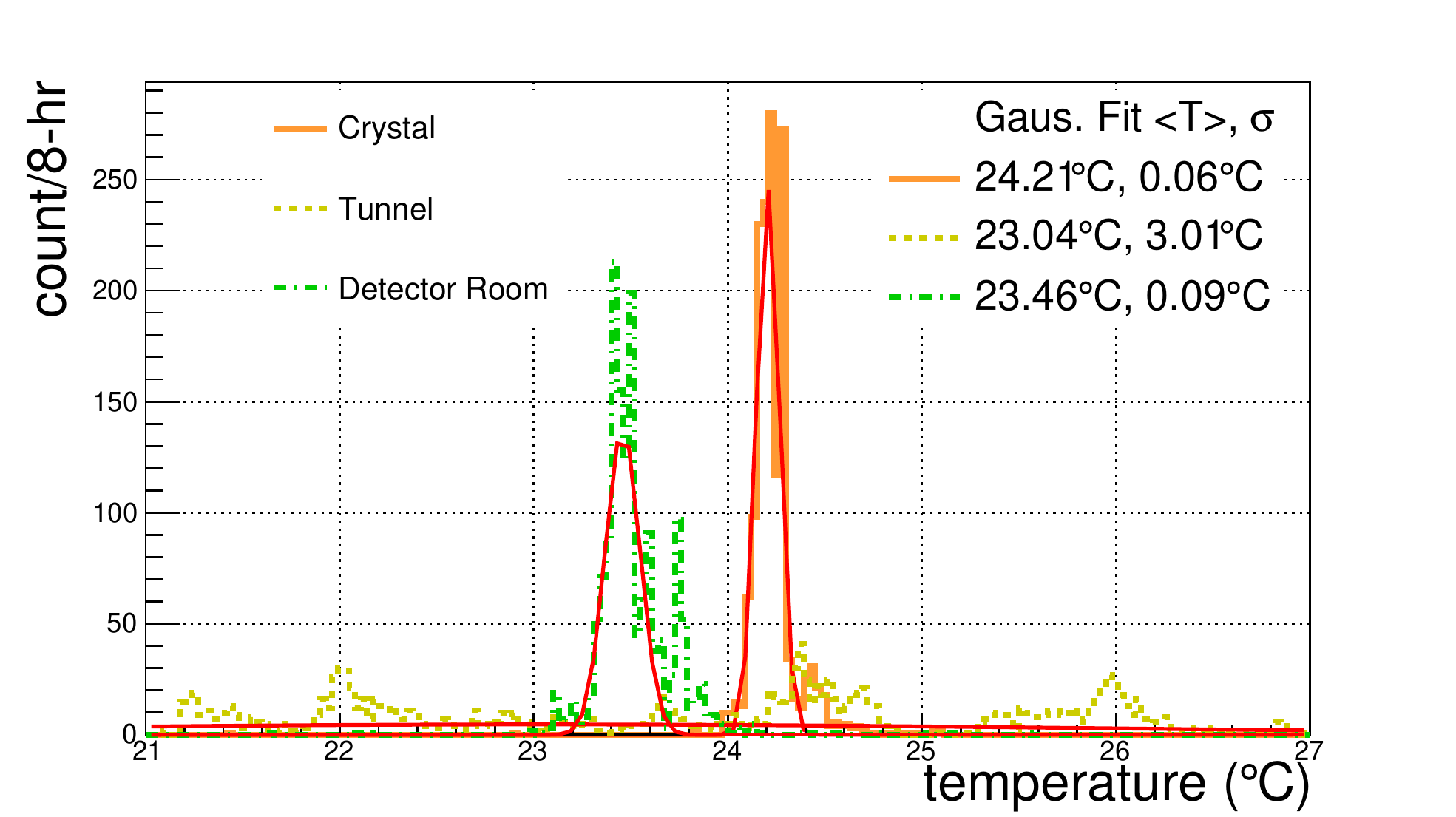}
    \caption{The long-term variation of the temperatures.
      The Gaussian fits are overlaid with thin red lines.
      The solid orange line indicates the crystal temperature variation over 1.5 years
      since the beginning of the data taking.
    }
    \label{tempvar}
  \end{center}
\end{figure}

The temperature in the detector room is well controlled to within a 0.09~$^{\circ}$C standard deviation, while the
detector (LS) temperature only varies by 0.06~$^{\circ}$C as shown in Fig.~\ref{tempvar}. The high voltages applied to PMTs do not vary by more than 1~V from their set value.
The humidity in the detector room varies by $\pm$1.0\,\%RH throughout a year while
humidity at the top of the liquid scintillator is measured to be 1.7\,\%RH\footnote{The standard limit of errors provided by the producer is 5\,\%RH.}.
The average radon level is stable at 36.7~Bq/m$^3$ with its standard deviation of $\pm$5.5 Bq/m$^3$.
Additional parameters including electricity, dust level, and pre-amplifier power
show no significant variations over the year.
The overall environmental conditions in the detector room are well monitored and very stable.
The measured long-term stability is summarized in Table~\ref{tab}.

\begin{table*}[!htb]
  \begin{center}
    \caption{Summary of environmental stability measured at least for 1.5 years.}
    \label{tab}
    \begin{tabular}{lcccc}
      \hline
      \hline
      Variable &  average values & standard deviations & location &  \\
      \hline
      Temperature &  23.46 $^{\circ}$C & 0.09$^{\circ}$C  & detector room &  \\
                  &  24.21 $^{\circ}$C & 0.06$^{\circ}$C  & crystal &  \\
                  &  23.04 $^{\circ}$C & 3.01$^{\circ}$C  & tunnel &  \\
      \hline
      Humidity    &  36.7\,\%RH &  1.0\,\%RH  & detector room &  \\
                  &  1.7\,\%RH  &  0.1\,\%RH & liquid scintillator top &  \\
      \hline
      Radon       &  36.7 Bq/m$^3$&5.5 Bq/m$^3$  & detector room &  \\
      \hline
      \hline
      
    \end{tabular}
  \end{center}
\end{table*}

\section{Conclusion}

The environmental monitoring system is critical in the dark matter direct experiment
such as the COSINE-100 experiment which aims at replicating the DAMA/LIBRA experiment results that show
a 1\,\% annual variation signal.
We have monitored various environmental parameters throughout the detector operation
that might induce artificial time-dependent modulation data.
The temperatures and humidities are controlled below the 1\,\% level.
The variation of the radon level in the room is relatively large but
it does not show any correlation with the experimental data.
With a stable monitoring system, we not only monitor the detector stability
but also provide safety measures.
We would like to apply further a similar concept with an improved design to the upcoming underground laboratory Yemilab. Modular subsystem in each experimental hole would include its monitoring platform with scalability.
For example, central tunnel monitoring parameters such as electricity, air circulation and
radon reduction would be provided as common monitoring values in this platform.
A specific experiment can add on its monitoring parameters separately.

\acknowledgments
We thank the Korea Hydro and Nuclear Power(KHNP) Company for providing underground laboratory space at Yangyang. This work is supported by:the Institute for Basic Science (IBS) under project code IBS-R016-A1 and NRF-2016R1A2B3008343, Republic of Korea; NSF Grants No. PHY-1913742, DGE1122492, WIPAC, the Wisconsin Alumni Research Foundation, United States; STFC Grant ST/N000277/1 and ST/K001337/1, United Kingdom; Grant No. 2017/02952-0 FAPESP, CAPES Finance Code 001, CNPq 131152/2020-3, Brazil.

\end{document}